# Мощные оптоэлектронные коммутаторы нано- и пикосекундного диапазона на основе высоковольтных кремниевых структур с *p-n*-переходами. II. Энергетическая эффективность.

## А. С. Кюрегян

Всероссийский Электротехнический институт им. В. И. Ленина, 111250, Москва, Россия

Впервые изучена энергетическая эффективность оптоэлектронных коммутаторов на основе высоковольтных кремниевых фотодиодов, фототранзисторов и фототиристоров, управляемых пикосекундными лазерными импульсами, при формировании импульсов напряжения на активной нагрузке $R_L$. Показано, что при заданных значениях амплитуды $U_R$ и длительности $t_R$ импульсов существуют оптимальные величины площади приборов, энергии и коэффициента поглощения управляющего излучения, обеспечивающие максимальный общий коэффициент полезного действия коммутатора около 0,92. Все три типа коммутаторов обладают практически одинаковой эффективностью при малых $t_R$, а при больших $t_R$ заметным преимуществом обладают фототиристоры.

В работе автора [1] были изложены результаты численного моделирования переключения высоковольтных кремниевых структур с *p-n*-переходами, управляемых пикосекундными лазерными импульсами, и получены соотношения между параметрами коммутаторов и характеристиками процесса переключения. Настоящая статья является продолжением [1], однако теперь основное внимание будет сосредоточено на анализе не изученного ранее вопроса об энергетической эффективности оптоэлектронных коммутаторов, которая фактически и определяет целесообразность практического применения этих приборов.

Все обозначения, а также объекты и метод исследования остались теми же, как и значения основных параметров коммутаторов: $U_0 = 5$ kV, $t_{ph} = 10$ ps, $R_L = 5$ Ohm, $T = 75$ C и, если это особо не оговорено, $S_{ph} = 0.5$ cm$^2$, $\kappa = 32$ cm$^{-1}$

Энергетическая эффективность коммутаторов характеризуется общим коэффициентом полезного действия (КПД)

$$\eta_{tot} = \frac{W_R}{W_R + W_{tot}}, \quad (1)$$

где $W_R$ - энергия, рассеянная сопротивлением нагрузки $R_L$, $W_{tot} = W_D + W_{ph}/\eta_{las}$ - общая энергия потерь коммутатора, $W_D$ энергия рассеянная переключающим прибором, $W_{ph}$ - энергия управляющего импульса света, $\eta_{las}$ - КПД лазера «от розетки», который мы полагали равным 0,1 [2]. Все эти энергии зависят от времени, как изображено на **Рис. 1**. Поэтому КПД коммутатора должен зависеть от длительности импульса $t_R$.

На первом этапе процесса коммутации емкость структуры $C_D$ разряжается током проводимости. Энергия, рассеиваемая за это время нагрузкой пренебрежимо мала, поэтому практически вся энергия $W_C = C_D U_0^2/2 \approx 0.24$ mJ, накопленная в барьерной емкости прибора, рассеивается им самим.

На втором этапе мощность, рассеиваемая фотодиодом пренебрежимо мала, так что энергия $W_D$ остается практически постоянной и равной $W_C$, а $W_R$ линейно увеличивается со временем вплоть до начала формирования ОПЗ в фотодиоде. в момент $t = t_{sc}$. Далее напряжение на фото-

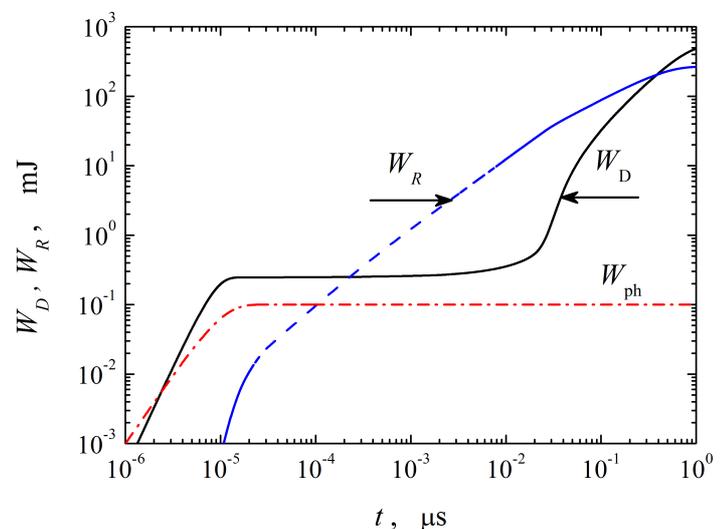

Рис. 1. Зависимости энергий, рассеиваемых нагрузкой $W_R$ и фотодиодом $W_D$ от времени при $W_{ph} = 100$ μJ.

диоде и, следовательно, энергия $W_D$ увеличиваются значительно быстрее $W_R$. Все это приводит к немонотонным зависимостям $\eta_{tot}$ от $t_R$, примеры которых изображены на

Рис. 2. Видно, что функция $\eta_{tot}(t_R)$ достигает своего максимума $\eta_{max}$ при некотором значении $t = t_{max}$.

С ростом $W_{ph}$ максимальный КПД $\eta_{max}$ сначала быстро увеличивается, но при больших $W_{ph}$ стремится к постоянной величине порядка 0,95 (см. Рис. 3). В то же время энергия $W_{Rmax} = W_R(t_{max})$, рассеиваемая нагрузкой за время $t_{max}$, и само время $t_{max}$ линейно увеличиваются с ростом $W_{ph}$, в частности

$$W_{Rmax} \approx 300 W_{ph}. \qquad (2)$$

Практически такие же зависимости $W_{Rmax}$, $t_{max}$ и $\eta_{max}$ от $W_{ph}$ получились для фототранзисторов во всем диапазоне значений $W_{ph}$, а для фототиристоров только при $W_{ph} \leq 35$ μJ, когда еще существует локальный максимум функции $\eta_{tot}(t_R)$. Величины $\eta_{max}$ и $t_{max}$ зависят также от площади приборов и от коэффициента поглощения света, как это изображено на Рис. 4 и Рис. 5.

Из рисунков 3-6 следует, что практически во всех рассмотренных случаях $t_{max} < t_{av}$. Это означает, что падение КПД ограничивает допустимую длительность импульса, формируемого фотодиодным коммутатором сильнее, чем наступление опасного динамического лавинного пробоя.

Если длительность $t_R$ импульса напряжения считать априори заданной величиной наряду с амплитудой $U_R \approx U_0/2$ и $R_L$, то остается три свободных параметров коммутатора: $W_{ph}$, $S_{ph}$ и $\kappa$. Зависимости $\eta_{tot}$ и $W_D$ от этих параметров изображены на Рис. 6-8 для $t_R = 10$ нс. Видно, что существуют оптимальные значения $W_{ph} \approx 50$ μJ, $S_{ph} \approx 0.5$ cm$^2$ и $\kappa \approx 32$ cm$^{-1}$, при которых КПД $\eta_{tot}$ максимален, а энергия $W_D$ и/или $W_{tot}$ минимальна. Причина существования минимума функции $W_{tot}(W_{ph})$ очевидна – см. Рис. 6.

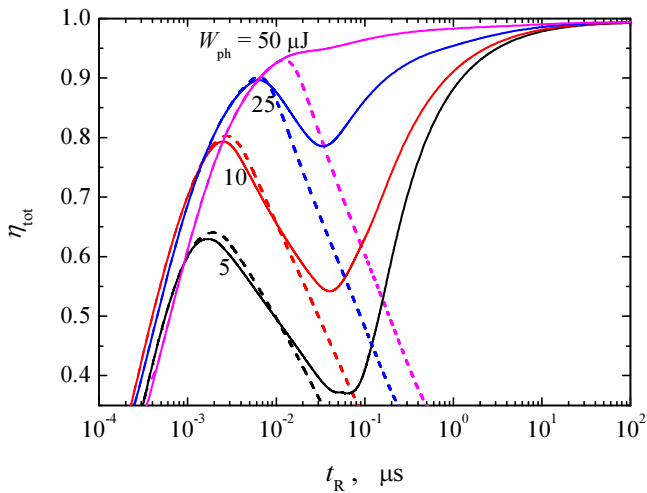

Рис. 2. Зависимость общего КПД $\eta_{tot}$ фотодиодного (штриховые линии) и фототиристорного (сплошные линии) коммутаторов от длительности импульса тока через нагрузку $t_R$ при $\eta_{las} = 0.1$ и различных значениях $W_{ph}$.

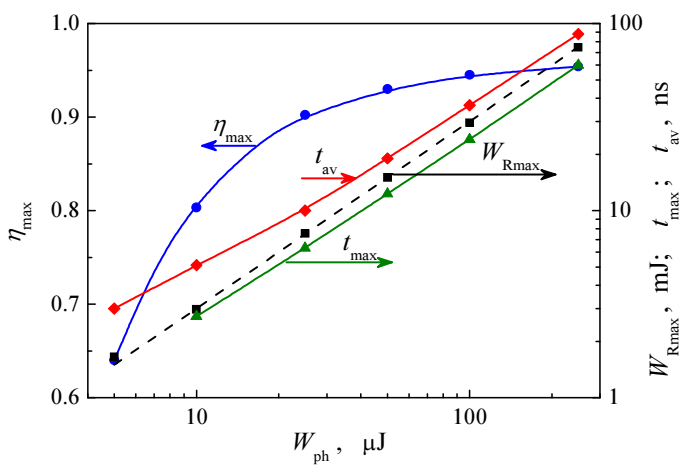

Рис. 3. Зависимости максимального полного КПД $\eta_{max}$ фотодиодного коммутатора, энергии $W_{Rmax}$, рассеиваемой нагрузкой за время $t_{max}$, и времен $t_{max}$, $t_{av}$ от $W_{ph}$ при $\eta_{las} = 0.1$. Штриховая линия – расчет по формуле (2).

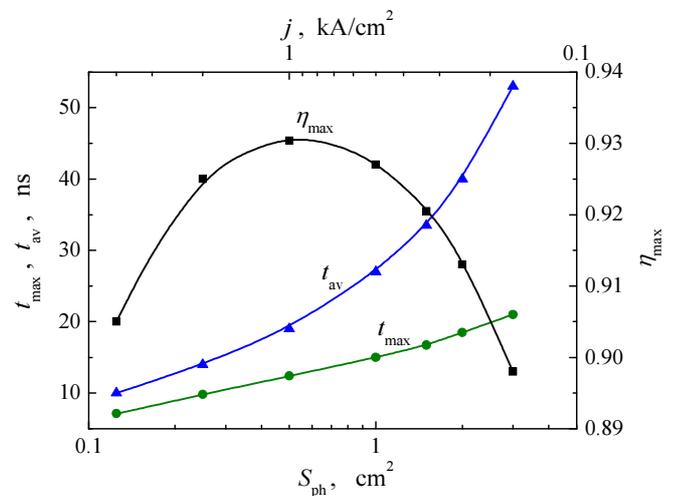

Рис. 4. Зависимости максимального полного КПД $\eta_{max}$ фотодиодного коммутатора и времени $t_{max}$ от $S_{ph}$ при $W_{ph} = 50$ μJ.

С ростом $S_{ph}$ уменьшается плотность тока, поэтому увеличивается время $t_{sc}$ возникновения ОПЗ (см. Рис. 7 в [1]), и, следовательно, снижаются потери на срезе импульса. В то же время потери на фронте, равные $W_C$, пропорциональны $S_{ph}$, что и приводит к возникновению минимума функции $W_{tot}(S_{ph})$.

Наконец, минимум функции $W_D(\kappa)$, изображенной на Рис. 8, существует потому, что при малых $\kappa$ концентрация неравновесных носителей заряда мала во всей структуре, а при больших – в прианодной области [1], где начинается «преждевременное» формирование ОПЗ и рост потерь на срезе импульса.

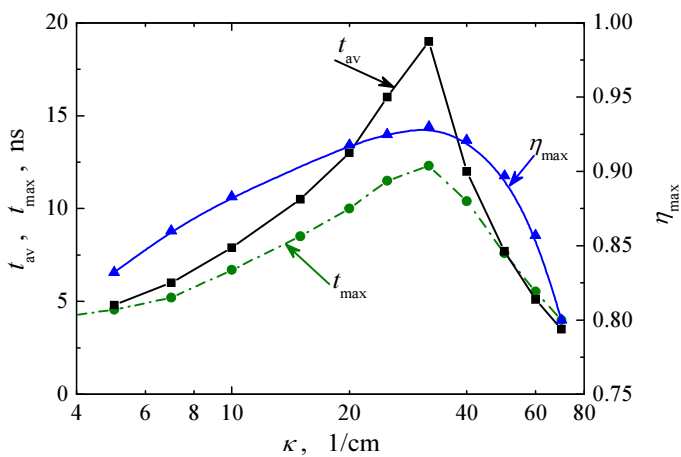

Рис. 5. Зависимости максимального полного КПД $\eta_{max}$ фотодиодного коммутатора и времен $t_{max}, t_{av}$ от коэффициента поглощения $\kappa$ для фотодиода при $W_{ph} = 50$ μJ.

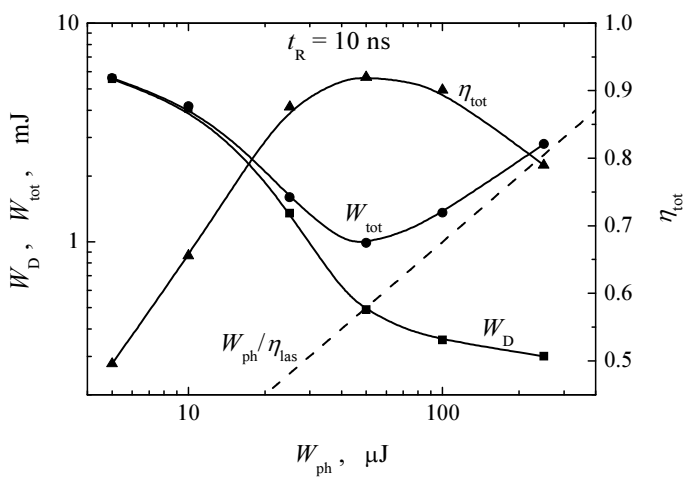

Рис. 6. Зависимости энергий $W_D, W_{tot}$ и общего КПД фотодиодного коммутатора $\eta_{tot}$ от $W_{ph}$ при длительности импульса тока нагрузки $t_R = 10$ ns.

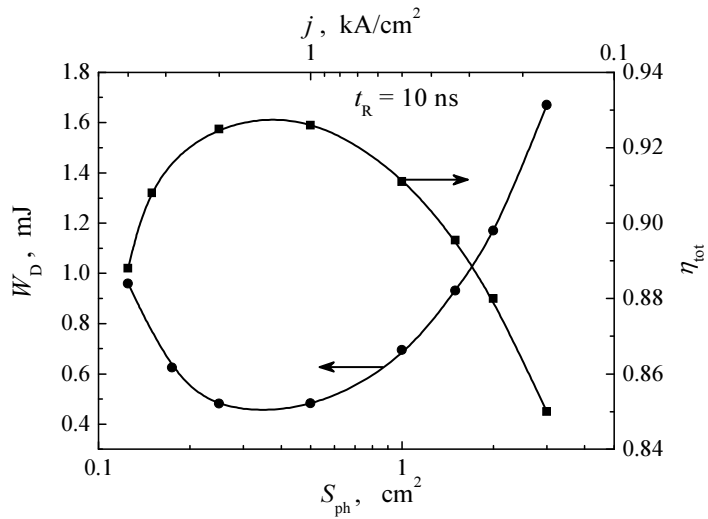

Рис. 7. Зависимости энергии $W_D$ и общего КПД фотодиодного коммутатора $\eta_{tot}$ от $S_{ph}$ при длительности импульса тока нагрузки $t_R = 10$ ns.

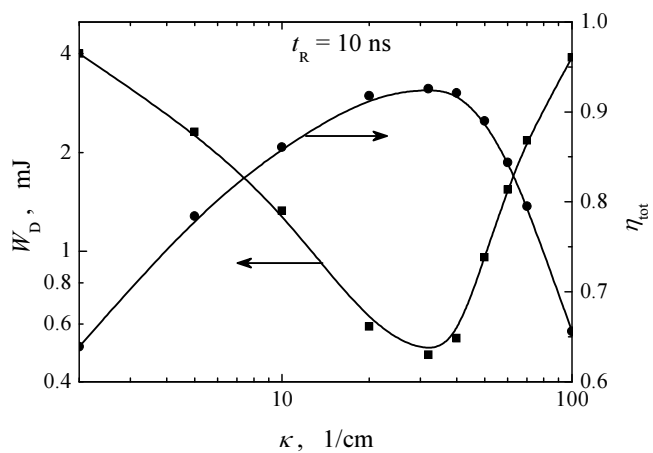

Рис. 8. Зависимости энергии $W_D$ и общего КПД коммутаторов $\eta_{tot}$ от $\kappa$ при длительности импульса тока нагрузки $t_R = 10$ ns.

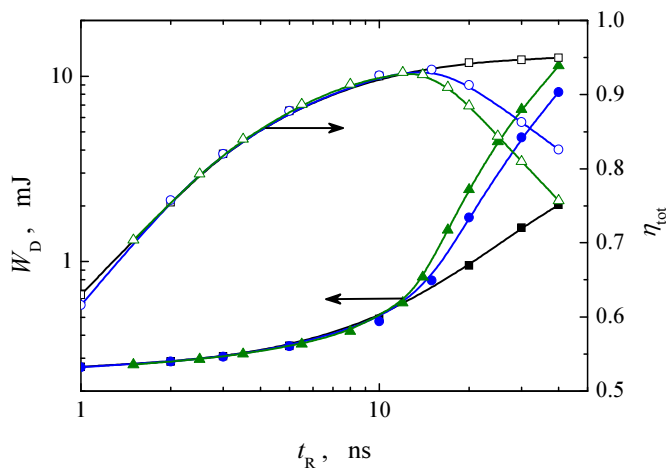

Рис. 9. Зависимости энергии $W_D$ и общего КПД $\eta_{tot}$ фотодиодных (треугольники), фототранзисторных (кружки) и фототиристорных (квадраты) коммутаторов от длительности импульса $t_R$ при $W_{ph} = 50$ μJ.

Приведенные на рисунках 3-8 результаты моделирования относятся к фотодиодным коммутатором. Однако, как было показано в [1], характеристики процессов переключения всех трех приборах практически идентичны до тех пор, пока в момент $t_{sc}$ не начинают формироваться области пространственного заряда. Только после этого начинает проявляться обычная инжекция одного (в фототранзисторах) или обоих (в фототиристорах) типов носителей заряда, отсутствующая в фотодиодах. Естественно, что и энергетические характеристики всех трех приборов должны совпадать при $t_R < t_{sc}$. Анализ результатов моделирования показывает, что они совпадают даже вплоть до $t_R \sim t_{max} \sim (1.5-2)t_{sc}$. Например, при оптимальных значениях $W_{ph} \approx 50$ µJ, $S_{ph} \approx 0.5$ cm$^2$ и $\kappa \approx 32$ cm$^{-1}$ преимущества фототранзисторов и, особенно, фототиристоров начинает проявляться только при $t_R > t_{max} \approx 12.5$ ns (см. Рис. 9).

Полученные результаты указывают на то, что использование оптопары на основе волоконного лазера и высоковольтной кремниевой структуры позволит создать почти идеальный коммутатор, способный формировать многокиловольтовые импульсы напряжения с фронтом менее 100 пс, длительностью до 10 нс и частотой повторения десятки килогерц при общем КПД «от розетки» более 0,92. По совокупности этих параметров такие приборы должны значительно превосходить все известные автору коммутаторы нано- и пикосекундного диапазона.

**Литература**